\documentclass[12pt]{article}
\usepackage{graphicx}
\usepackage{multirow}
\usepackage{subfig}
\pdfoutput=1

\newcommand\pubdate{\today}

\def\fermi{Fermilab\\ Batavia IL, USA }

\def\Title#1{\begin{center} {\Large #1 } \end{center}}
\def\Author#1{\begin{center}{ \sc #1} \end{center}}
\def\Address#1{\begin{center}{ \it #1} \end{center}}

\newcommand\pubblock{\rightline{\begin{tabular}{l}\\
        \pubdate  \end{tabular}}}
\newenvironment{Abstract}{\begin{quotation}  }{\end{quotation}}
\newenvironment{Presented}{\begin{quotation} \begin{center} 
             PRESENTED AT\end{center}\bigskip 
      \begin{center}\begin{large}}{\end{large}\end{center} \end{quotation}}

\begin{document}
\begin{titlepage}
\pubblock

\vfill
\Title{Searches for Exotic Phenomena at ATLAS and CMS}
\vfill
\Author{ Sho Maruyama on behalf of the ATLAS and CMS Collaborations}
\Address{\fermi}
\vfill
\begin{Abstract}

The Large Hadron Collider (LHC) was operated at a center-of-mass energy of $\sqrt{s}$ = 7 and 8TeV for proton-proton collisions in Run I. 
The CMS and ATLAS detectors both collected approximately 20 $fb^{-1}$ of 8TeV data in the data taking period. 
This large  dataset collected at an unprecedented energy provides an ideal opportunity to search for new physics. 
In this paper, a selection of recent results from the ATLAS and CMS Collaborations concerning searches for exotic phenomena are presented. 
The signal models of these analyses contain heavy resonances, dark matter particles, and long-lived particles.

\end{Abstract}
\vfill
\begin{Presented}
XXXIV Physics in Collision Symposium \\
Bloomington, Indiana,  September 16--20, 2014
\end{Presented}
\vfill
\end{titlepage}
\def\thefootnote{\fnsymbol{footnote}}
\setcounter{footnote}{0}

\section{Physics motivation}
\label{sec:phys}
The particle content of the standard model (SM) is completed with the discovery of the SM-like Higgs boson at ATLAS and CMS~\cite{atlashiggs, cmshiggs}.
However, there are many questions still open. Just to name few;
ordinary matter accounting for $\sim$5\% of mass-energy in the universe and no dark matter candidate in the SM;
 Hierarchy problem \& fine tuning of SM Higgs mass;
 No explanation for fermion masses and mixings and three family structure;
 Unification of strong, electro-weak, and gravitational forces;
 Compositeness of leptons and quarks.

The most pressing question above is that of the dark matter. It is an experimental fact that there is something we cannot explain within the  SM, opposed to tuning of some parameters in the SM or beyond SM models.

\section{ Exotica searches}
Each exotica search tries to address one or more fundamental physics questions mentioned in Sec.~\ref{sec:phys}. 
Many exotica searches are signature based so that different models can be studied in the same final state. 
In such a signature, even never thought-of signals may show up unexpectedly.
Analyses presented in this paper are public results of the ATLAS and CMS experiments at CERN.
The details of ATLAS and CMS detectors can be found elsewhere~\cite{atlas, cms}.

Given the page limitations for this conference proceeding, only a subset of recent results are summarized.
Most of the results became public between the time of ICHEP 2014 and September 2014.
Other search results can be found at ~\cite{atlaspub, cmspub}.

\subsection{Dark matter particles}
\begin{figure}[htb]
\centering
\subfloat[ATLAS tensor (D9) exclusion \cite{PhysRevD.90.012004}]{\includegraphics[width=0.5\textwidth]{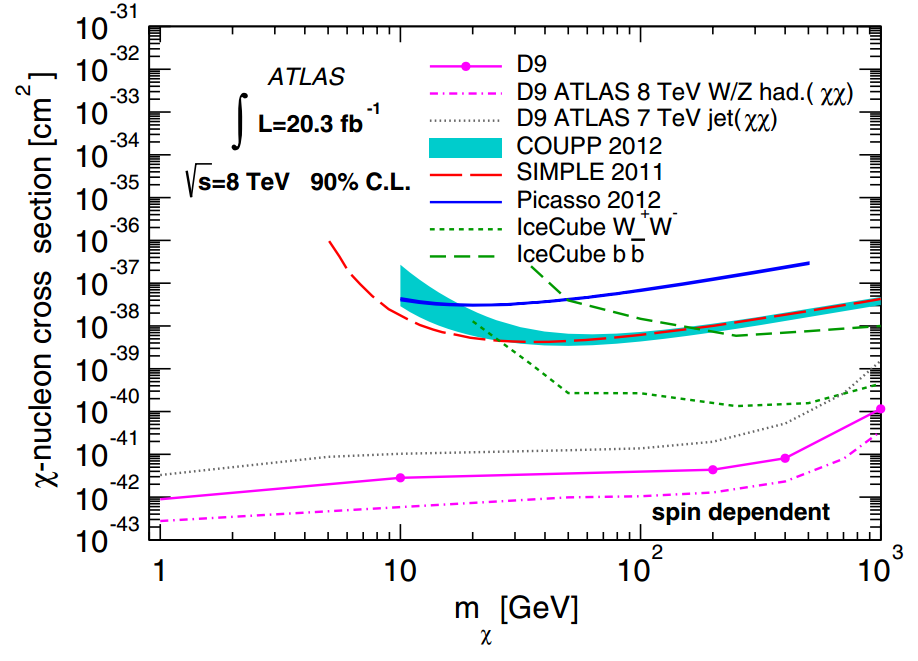}}
\subfloat[ATLAS scalar (D1) and vector (D5) exclusion~\cite{PhysRevD.90.012004}]{\includegraphics[width=0.5\textwidth]{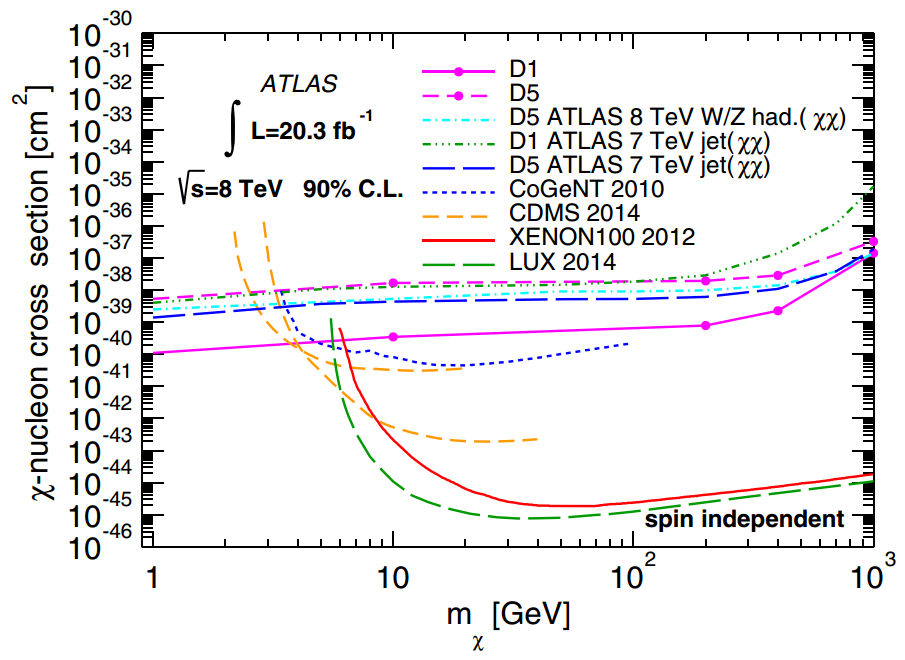}}
\hspace{0mm}
\subfloat[CMS axial--vector exclusion \cite{CMSPASEXO-12-047}]{\includegraphics[width=0.5\textwidth]{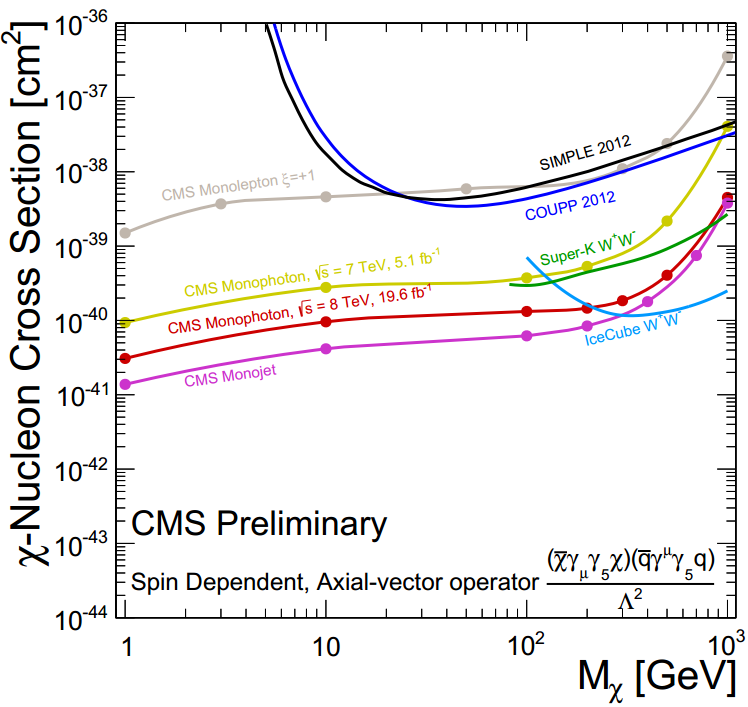}}
\subfloat[CMS vector exclusion \cite{CMSPASEXO-12-047}]{\includegraphics[width=0.5\textwidth]{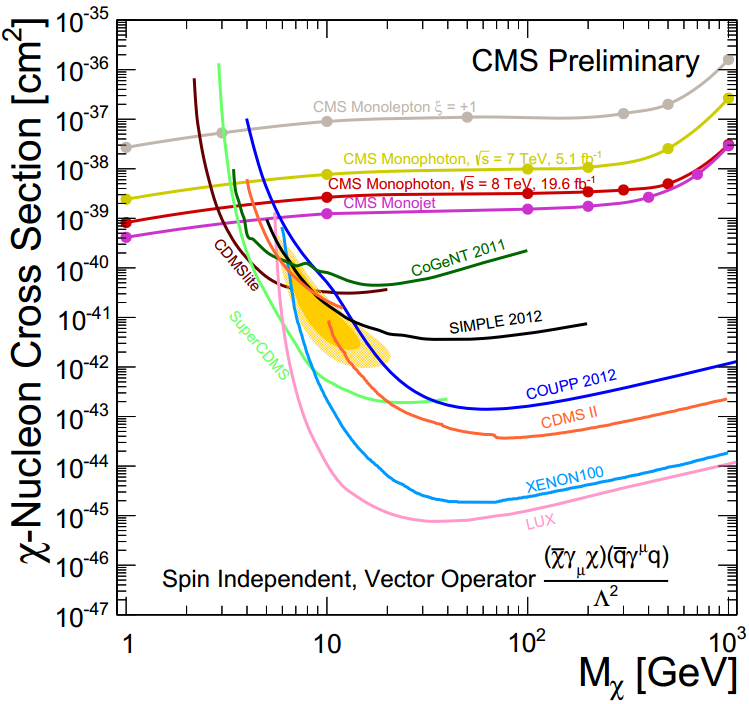}}
\caption{Dark matter search results.}
\label{fig:dmatlascms}
\end{figure}

Cold dark matter contributes approximately a quarter of the total mass-energy of the universe, while ordinary matter is only 4.9\% \cite{planck}. 
There is no SM particle to explain cold dark matter abundance, and thus there could be another particle which does not belong to the SM.
Cold dark matter relic density nicely matches to weakly interacting massive particles (WIMPs). 
Dark matter particle searches assume WIMP Dirac type effective field theory operators: scalar, vector, axial--vector, and tensor \cite{2010PhRvD..82k6010G, 2010JHEP...12..048B, 2011PhLB..695..185G}.  
At collider experiments, dark matter particles can be pair-produced with at least one initial state radiation, which provides sufficient ``kick'' in the transverse plane, leading to large missing transverse energy (MET).
Therefore searches are conducted in Mono--X+MET final states, where Mono--X can be a gluon, $\gamma$, or massive vector boson \cite{2011PhRvD..84i5013R, 2012PhRvD..85e6011F, 2013PhLB..723..384B}.
The ATLAS searches \cite{PhysRevD.90.012004,PhysRevLett.112.041802} are performed in leptonic Z + MET, and hadronic W or Z + MET final states. 
The CMS searches \cite{arXiv:1408.3583,CMSPASEXO-12-047} are performed in large transverse momentum ($p_{T}$) jet or $\gamma+$ MET final states.
The results are presented as exclusions in the mass of dark matter and nucleon-dark matter cross section for different types of operators in Fig.~\ref{fig:dmatlascms}.
An advantage of collider experiments is sensitivity in lower mass regions where direct searches lose sensitivity.

\subsection{W'}
\begin{figure}[htb]
\centering
\subfloat[ATLAS leptonic W' exclusion \cite{arXiv:1407.7494}]{\includegraphics[width=0.5\textwidth]{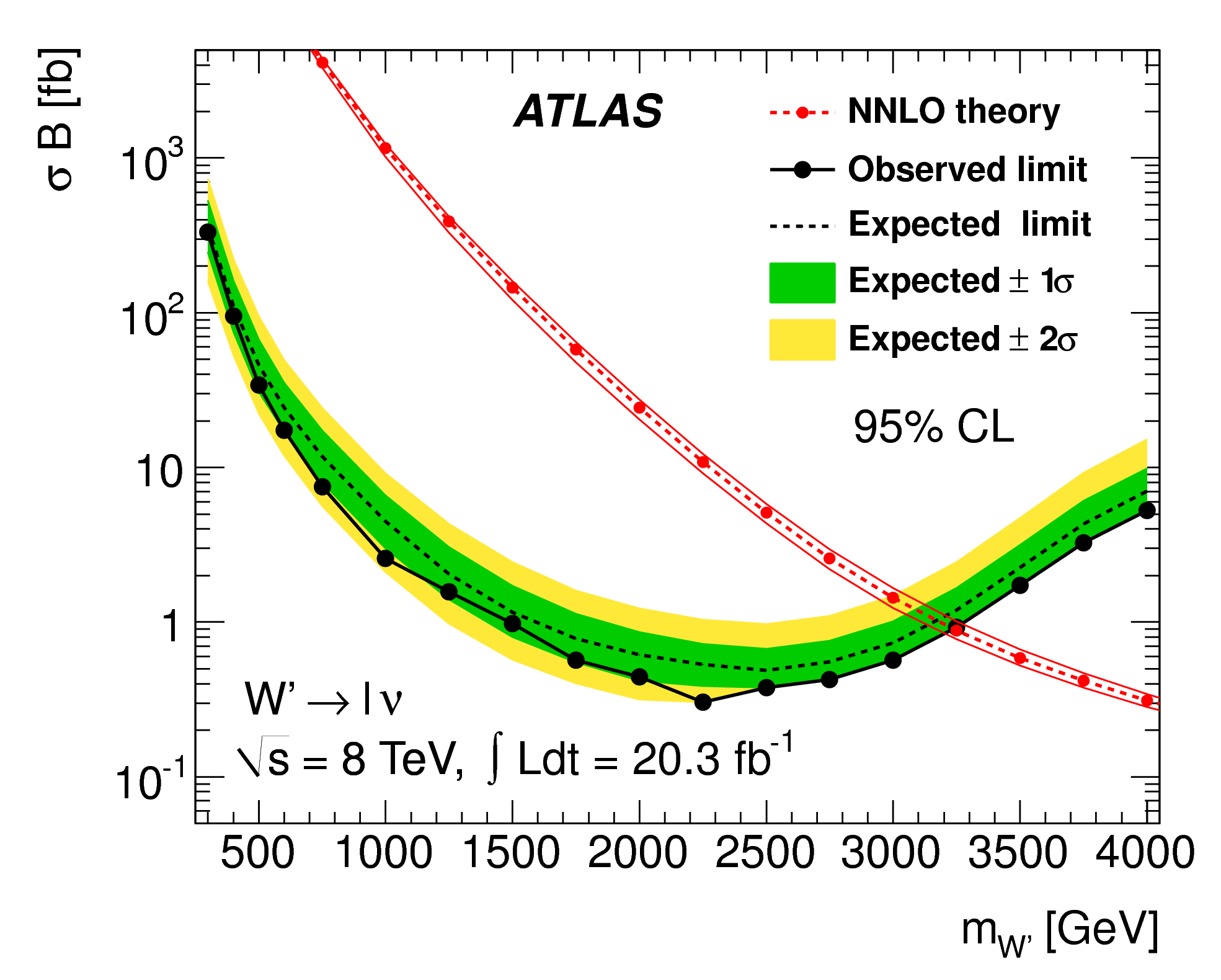}}
\subfloat[CMS leptonic W' exclusion \cite{arXiv:1408.2745}]{\includegraphics[width=0.5\textwidth]{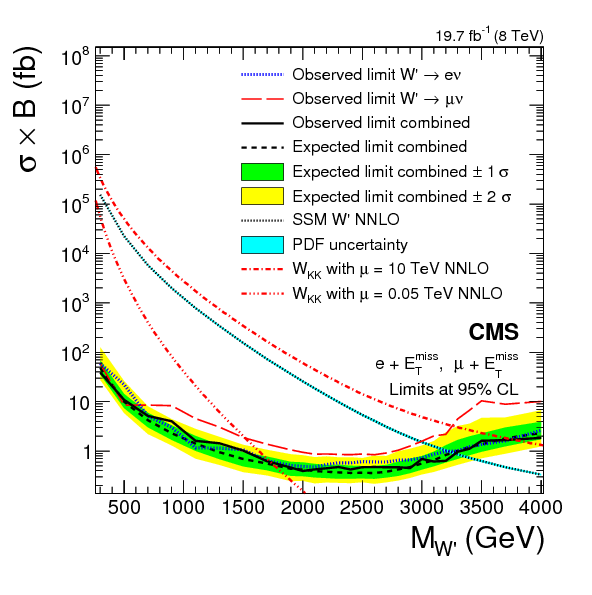}}
\caption{Leptonic W' search results.}
\label{fig:wpatlascms}
\end{figure}

The SM could be a remnant of a larger symmetry group, which unifies electro-weak and strong (and gravitational) forces. 
Heavy vector bosons arising from such broken symmetries include W' and Z' particles.
As a bench mark point, W' bosons are assumed to couple to fermions like SM W bosons do.
In the case of W' coupling to W and Z bosons, fermion decay modes are set to be negligible. 
It is important to search for W' bosons in different channels to account for a priori unknown W' couplings.
These searches overlap with the DM mono--W searches.
The ATLAS and CMS searches \cite{arXiv:1408.2745,arXiv:1407.7494} are performed in single lepton+MET, and hadronic W or Z + MET final states.
Obtained exclusion limits are compatible between the two experiments. 
The mass of W' boson below 3.28TeV is excluded at 95\% confidence level (CL) (see Fig.~\ref{fig:wpatlascms}). 
The ATLAS search \cite{arXiv:1408.0886} is performed in the boosted jets final state, where a W' boson decays to a top quark and b quark.
Unlike SM W bosons, this new diquark mode opens up because W' bosons are much heavier than SM W bosons.
For this particular final state, the mass of W' boson below 1.68TeV is excluded at 95\% CL.
The details of the jet substructure technique can be found in~\cite{arXiv:1011.2268}.

\subsection{Scalar lepto-quarks}
\begin{figure}[htb]
\centering
\subfloat[Fisrt generation scalar LQ exclusion \cite{CMS-PAS-EXO-12-041}]{\includegraphics[width=0.50\textwidth]{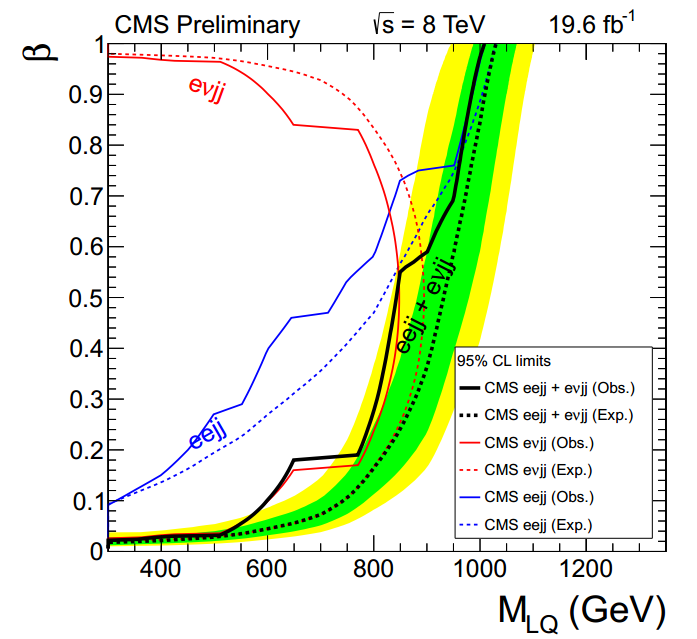}}
\subfloat[Third generation scalar LQ exclusion \cite{CMS-PAS-EXO-13-010}]{\includegraphics[width=0.50\textwidth]{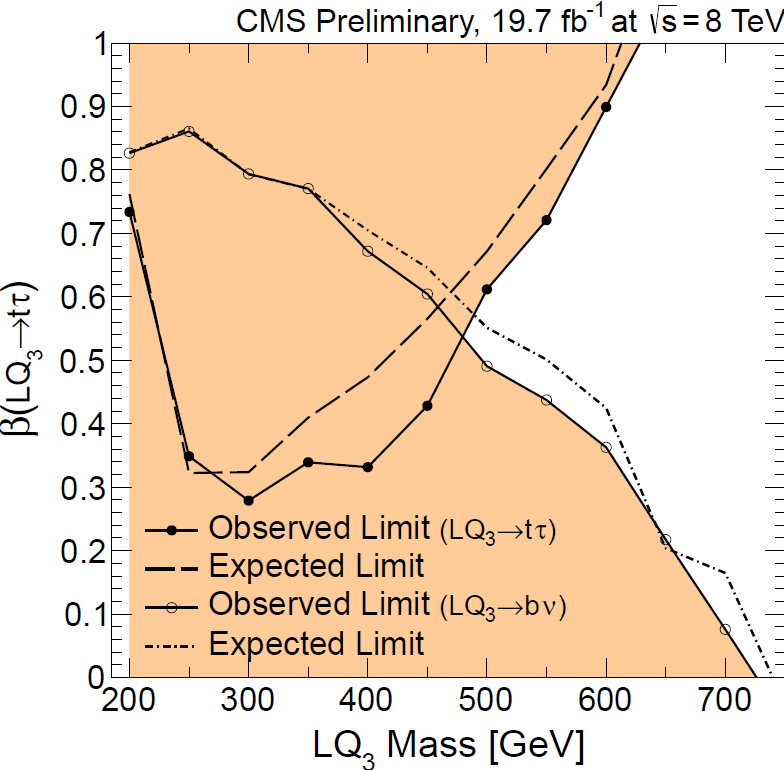}}
\caption{Scalar LQ search results.}
\label{fig:lqatlascms}
\end{figure}

The SM structure suggests a fundamental relationship between leptons and quarks. 
Lepto-quarks (LQs), which are scalar or vector bosons carrying color and fractional electric charges, can arise from a larger symmetry group such as SU(5) grand unification, SU(4) Pati-Salam, and $E_{6}$ models.
Measurements on flavor changing neutral current, lepton family number violation, and other rare decays favor that LQs decay within the same generation.
The decay modes of LQs are parametrized by $ \beta = BR(LQ \rightarrow \textnormal{charged lepton plus quark}) $. 
First and third generation scalar LQs searches are performed at CMS in final states of eejj and e$\nu$jj, and $\mu \tau+$jets and e$\tau+$jets \cite{CMS-PAS-EXO-12-041, CMS-PAS-EXO-13-010,arXiv:1408.0806}.
All searches require large scalar sum of $p_{T}$ to reduce SM background processes.
Exclusion limits are set in the LQ mass and $\beta$ plane (see Fig.~\ref{fig:lqatlascms}).

\subsection{Lepton flavor violation}
Charged lepton flavor number turns out to be conserved in the SM.
However there is no associated symmetry to protect it, and thus it is said to be accidental; some higher dimensional operators appear as suppressed at the energy scale of the SM.
On the other hand, neutral lepton flavor number is not conserved in neutrino oscillation (e.g., electron neutrinos can become muon neutrinos and so forth).
Lepton flavor violation also serves as a constraint on BSM models where lepton flavor violating terms may be present.
ATLAS performs a search in Z$\rightarrow$e$\mu$ final state, which is dominated by the Z$\rightarrow\tau\tau\rightarrow$e$\mu\nu\nu$ process \cite{arXiv:1408.5774}.
Because the number of background events is expected to be much larger than that of signal events, a shape analysis is conducted in the mass distribution of e and $\mu$.
Most of the systematic uncertainties are canceled by normalizing to the measured Z$\rightarrow$ee and Z$\rightarrow\mu\mu$ events.
The upper bound on BR(Z$\rightarrow$e$\mu$) is set to be less than $7.5 \times 10^{-7}$, which is a factor of two improvement from the previous direct bound set by the OPAL collaboration \cite{opal}.

\subsection{Compositeness}
\begin{figure}[htb]
\centering
\subfloat[ATLAS CI exclusion \cite{arXiv:1407.2410}]{\includegraphics[width=0.50\textwidth]{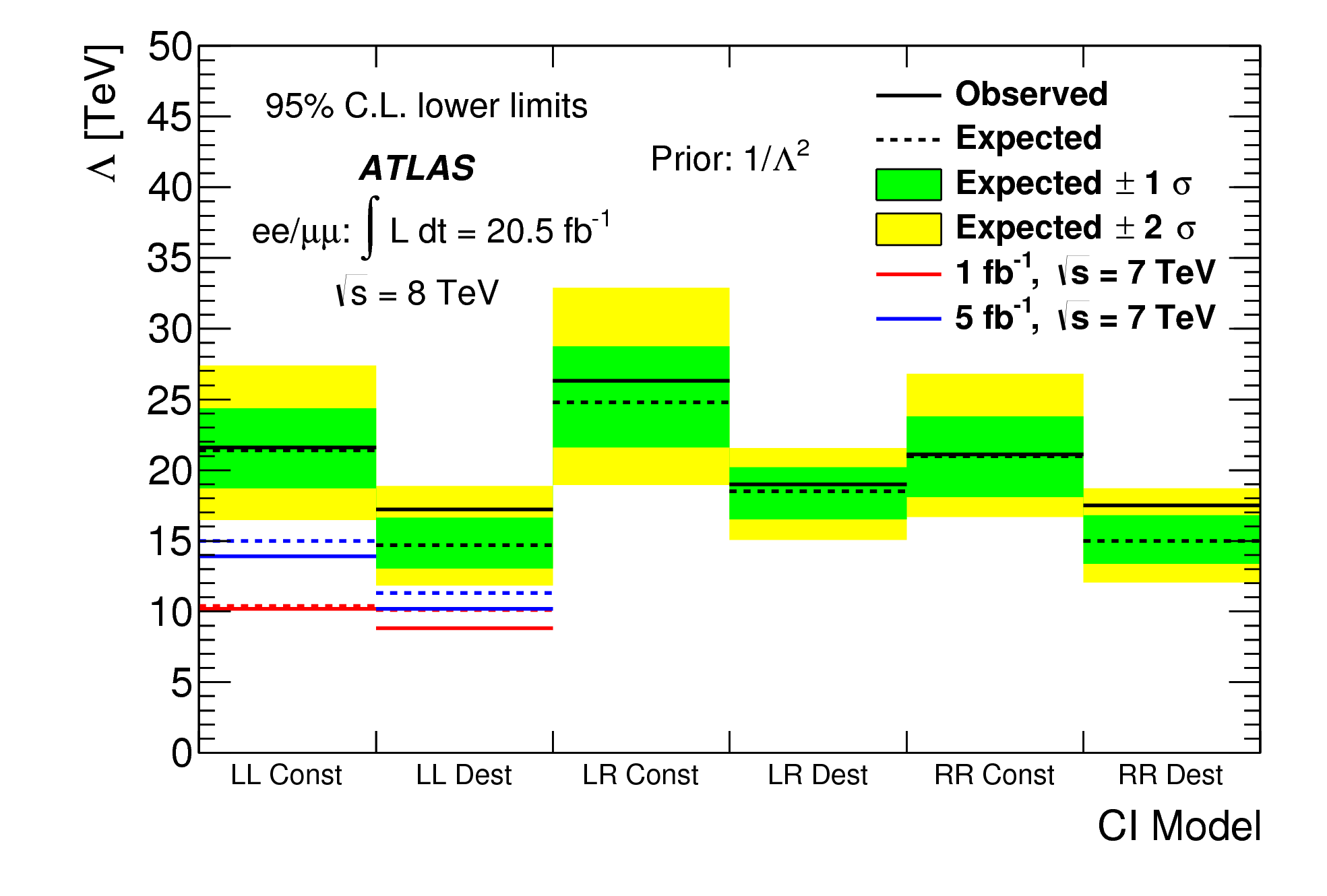}}
\subfloat[CMS CI exclusion \cite{CMS-PAS-EXO-12-020}]{\includegraphics[width=0.50\textwidth]{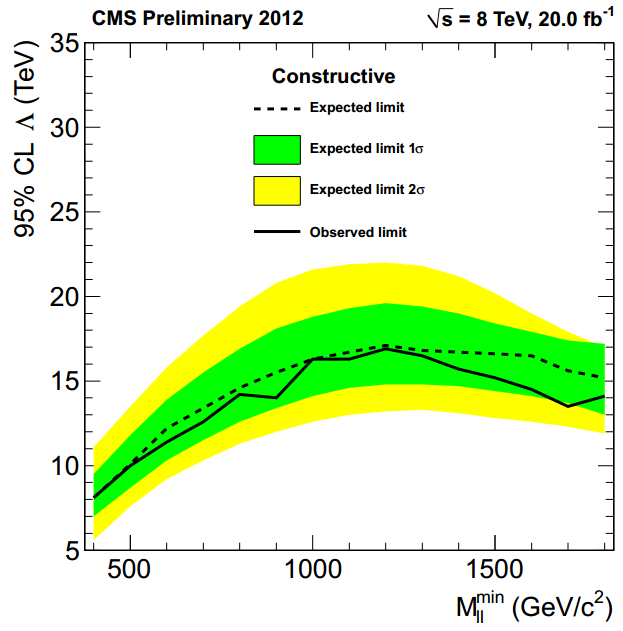}}
\caption{CI search results.}
\label{fig:ciatlascms}
\end{figure}

Leptons and quarks might be bound states of more fundamental particles, like atoms and nucleons were once considered as indivisible building block of matter.
ATLAS and CMS search for compositeness of leptons and quarks in two different models: contact interaction(CI) \cite{CMS-PAS-EXO-12-020,arXiv:1407.2410}  and excited quarks \cite{arXiv:1406.5171,arXiv:1407.1376}.
The former is analogous to Fermi's contact interaction theory. Even if the interacting energy scale is much higher than the electro-weak scale, still the effect of new physics can be detected at a lower energy scale, like weak force in $\beta$ decays.
Signals are sought in dilepton mass distribution where signal processes interfere with Drell-Yan processes.
The ATLAS search utilizes the angular distribution of leptons on top of the dilepton mass to improve sensitivity to signal events. 
Exclusion limits are set on the interaction energy scale for different handedness combinations and interference scenarios (see Fig.~\ref{fig:ciatlascms}).

\begin{figure}[htb]
\centering
\subfloat[ATLAS excited quark exclusion \cite{ arXiv:1407.1376}]{\includegraphics[width=0.50\textwidth]{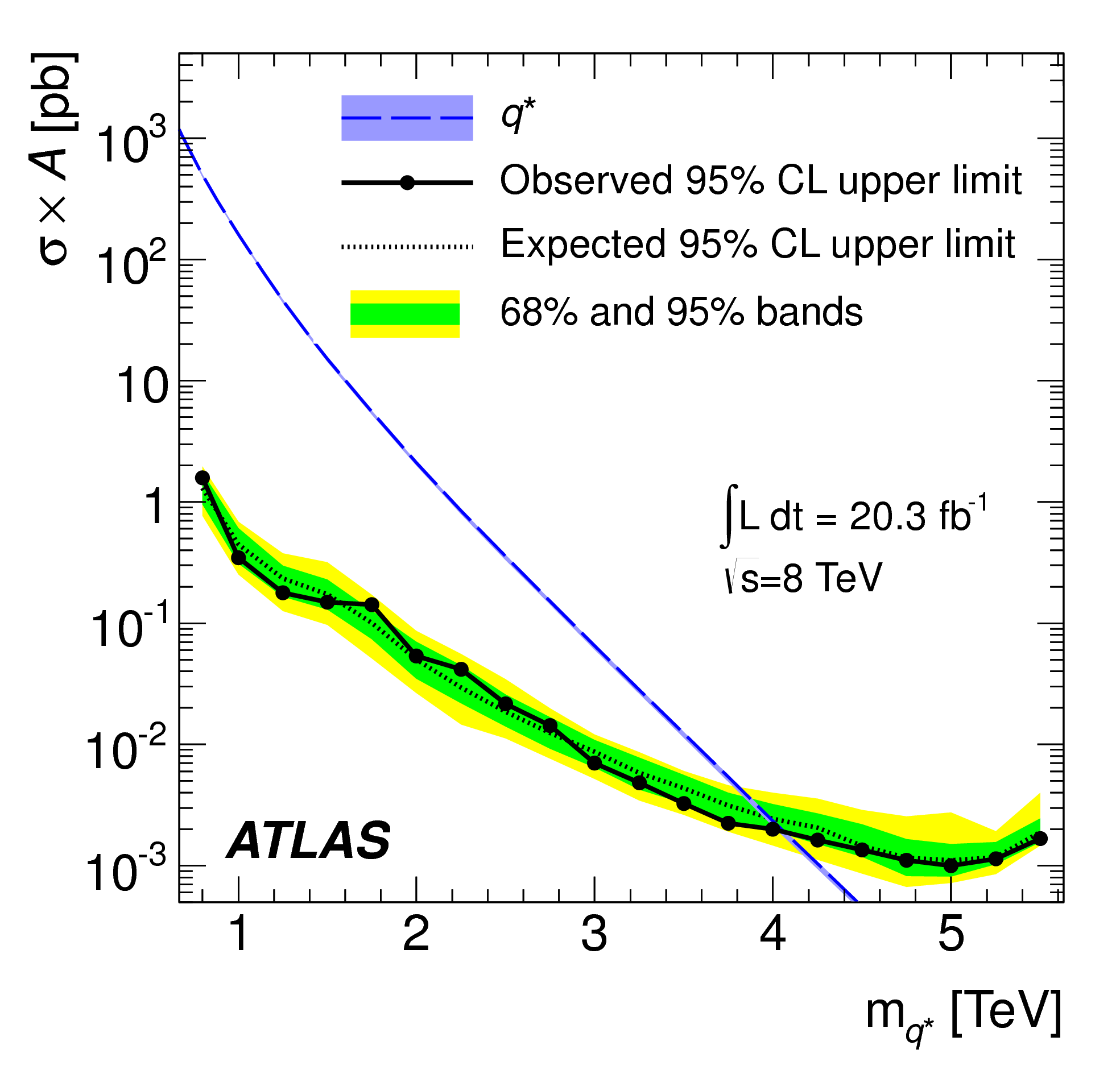}}
\subfloat[CMS excited quark exclusion \cite{arXiv:1406.5171}]{\includegraphics[width=0.50\textwidth]{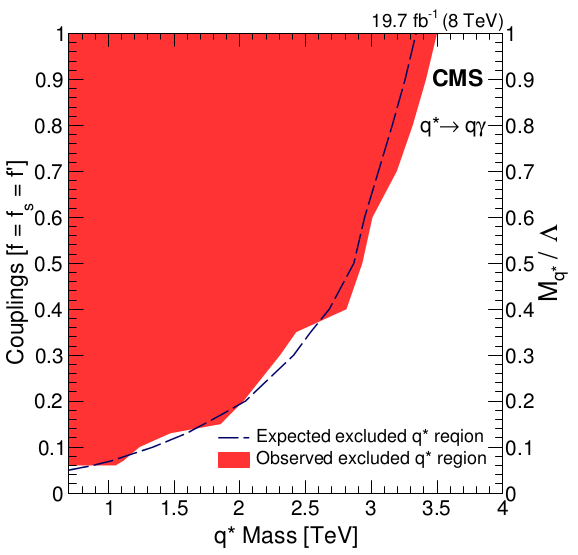}}
\caption{Excited quark search results.}
\label{fig:eqatlascms}
\end{figure}

The latter is analogous to excited hadrons where the bound states have larger orbital angular momenta.
Such excited bound states can transition down to ground states by radiation.
The ATLAS search is performed in the dijet final state, and the CMS search in $\gamma+$jet final state.
Exclusion limits on excited quark mass is set at 95\% CL by assuming SM-like couplings to fermions (see Fig.~\ref{fig:eqatlascms}).

\subsection{Vector-like quarks}
\begin{figure}[htb]
\centering
\subfloat[$T\bar{T}$ pair-production exclusion]{\includegraphics[width=0.50\textwidth]{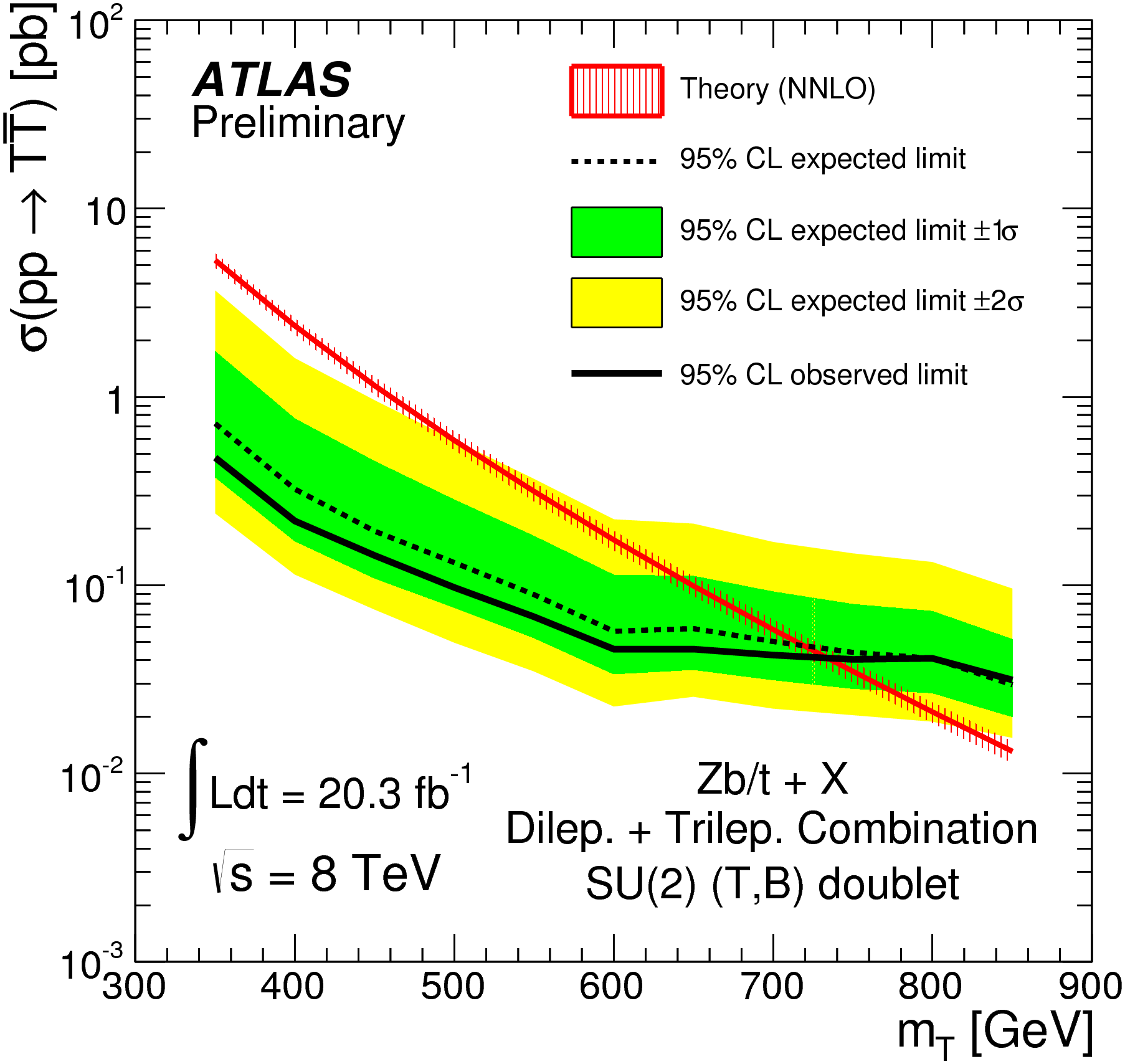}}
\subfloat[$B\bar{B}$ pair-production exclusion]{\includegraphics[width=0.50\textwidth]{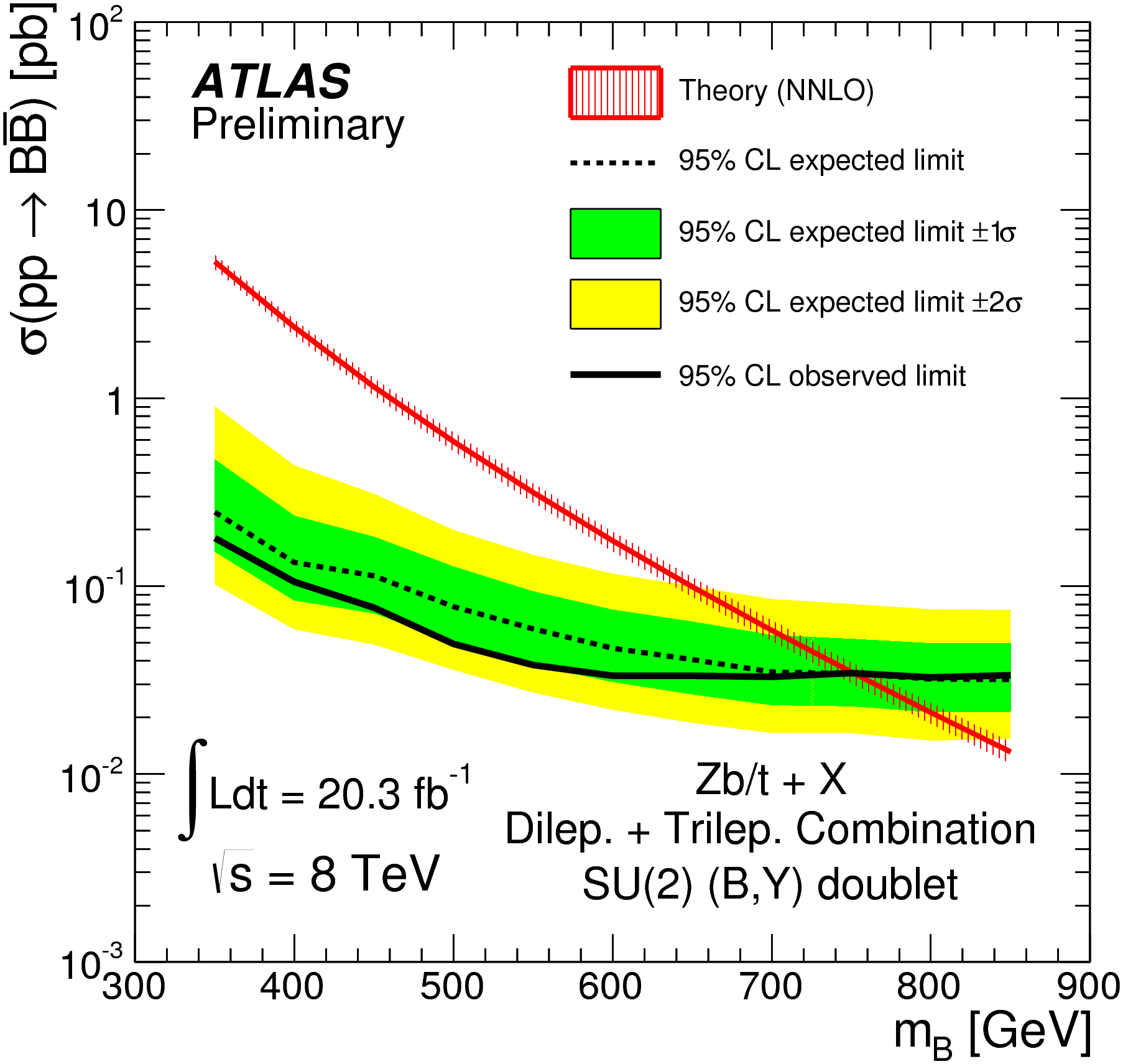}}
\caption{Vector-like quark search results \cite{ATLAS-CONF-2014-036}.}
\label{fig:vqatlascms}
\end{figure}

One of the unexplained features of the SM is why the Higgs mass is 125GeV. 
The radiative correction to the bare mass is quadratically divergent as well as independent of observed Higgs mass itself. 
One way to solve this problem is to nearly double the number of SM particles so that radiative corrections would cancel between SM particles and SUSY partners.
There are also non-SUSY ways to solve the problem. For instance, the Little Higgs model and Composite Higgs models have strongly coupled new states including
vector-like quarks. 
Such particles mix with SM quarks, leading to (partial) cancellation of radiative correction to Higgs mass.
Vector-like quarks (Q) can decay to a quark and massive boson. 
As a bench mark point, the branching fraction of $Q\rightarrow q+V$ is set to one third for each  boson (Higgs, Z, and W).
The ATLAS searches are performed for $T\rightarrow t+Z$ and $B\rightarrow b+Z$ in dilepton+jets and trilepton+jets final states \cite{ATLAS-CONF-2014-036}.
Exclusion limits are set on vector-like quark masses at 95\% CL (see Fig.~\ref{fig:vqatlascms}).

\subsection{Low scale technicolor}
Low scale technicolor models (LTC) can break electro-weak symmetry dynamically to generate fermion masses, and involve spin-0 and spin-1 particles. 
With the discovery of the SM-like Higgs boson, LTC on its own is inconsistent with experimental observations. 
However, LTC with a light composite Higgs boson is still valid, and it could be used as a bench mark model for a spin-1 resonance decaying to dibosons.
Techni-meson masses are chosen so that the techni-pion mode is kinetically inaccessible.
The ATLAS searches are performed in W$+\gamma$ and Z$+\gamma$ final states \cite{arXiv:1407.8150}, and the CMS search is performed in WZ trilepton final states \cite{arXiv:1407.3476}.
The mass ranges, [275,960], [200,700] and [750,890], and below 1140 GeV are excluded at 95\% CL for $a_{T}$, $\omega_{T}$, and $\rho_{T}$ techni-mesons respectively.

\subsection{Hidden valley}
\begin{figure}[htb]
\centering
\subfloat[ $\pi_{v}$ proper decay length exclusion \cite{ATLAS-CONF-2014-041}]{\includegraphics[width=0.50\textwidth]{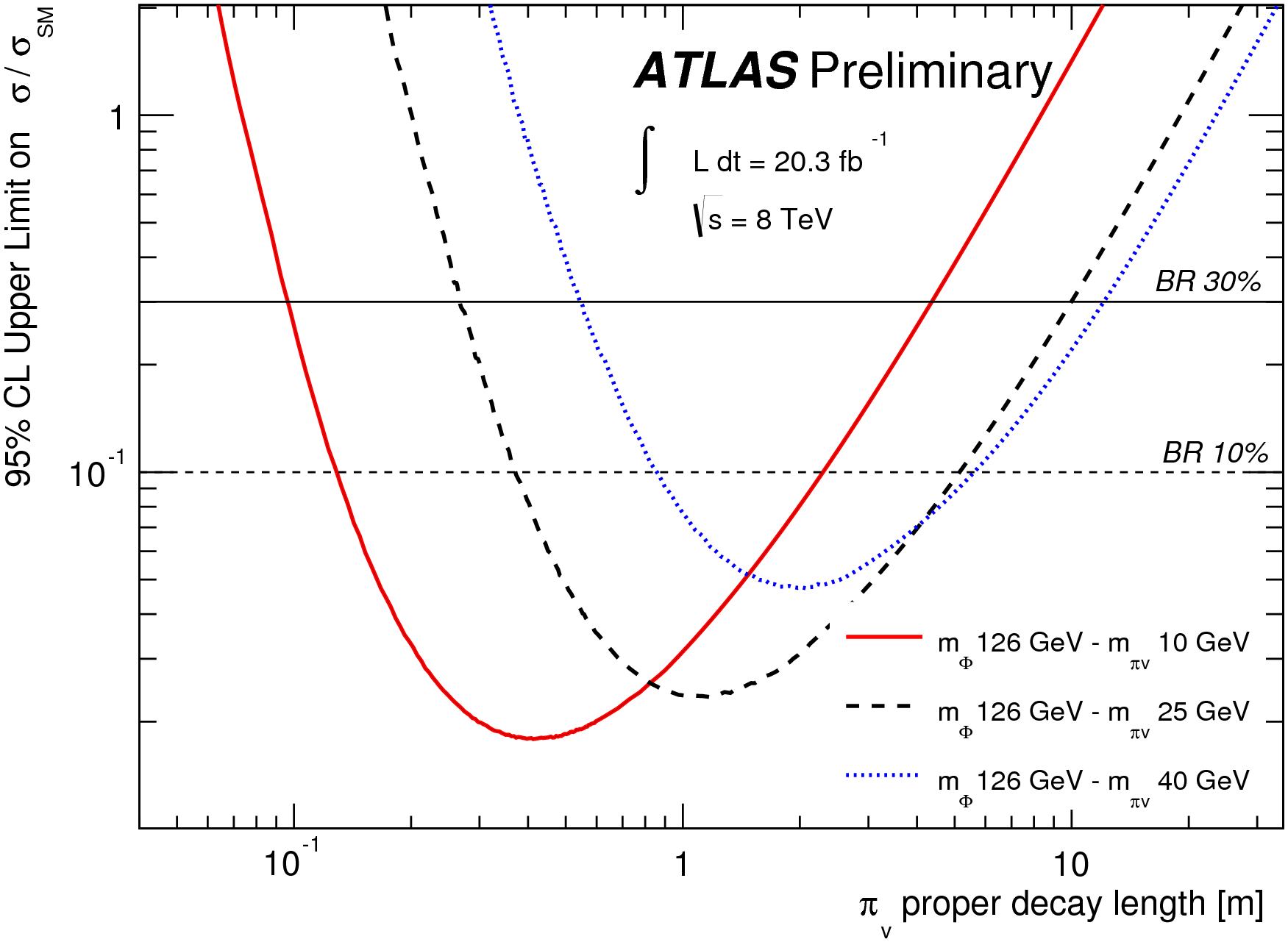}}
\subfloat[Dark photon exclusion \cite{arXiv:1409.0746}]{\includegraphics[width=0.50\textwidth]{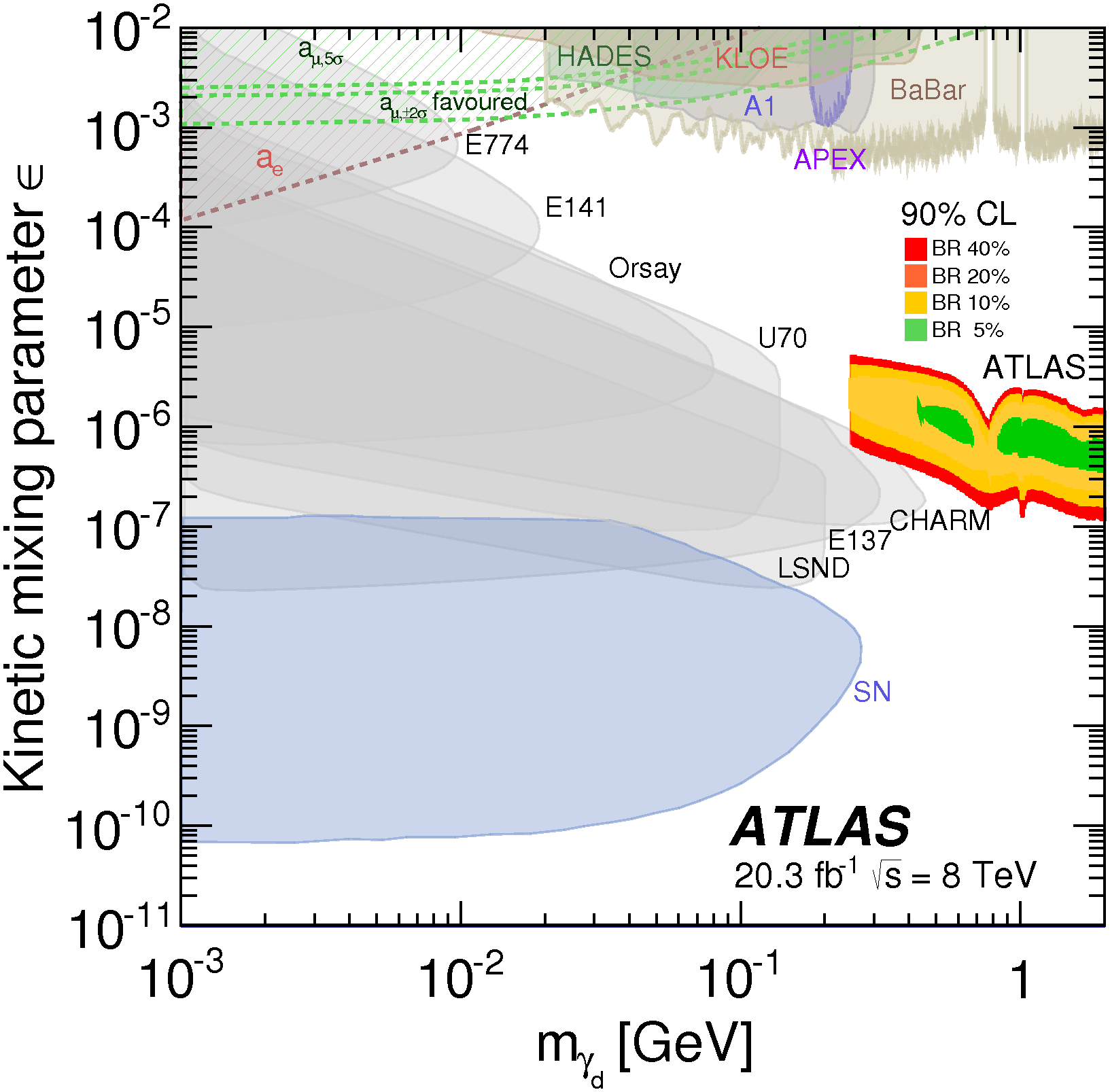}}
\caption{Hidden valley search results.}
\label{fig:hvatlascms}
\end{figure}

There could be a hidden sector (hidden valley), which is not gauged under the SM and thus not directly communicating with the SM sector. 
Such a sector can contain DM candidate particles (HLSPs), and long-lived particles which travel away from the beam interaction regions.
At some higher energy scales, suppose there are some operators which are charged under the SM and hidden sector. 
Through such heavy particles or small mixing with SM particles,  hidden particles can be produced at collider experiments.
Final states could be unique: displaced dilepton, displaced dijet, lepton jets, and collimated leptons. 
Background contributions can come from cosmic rays and beam halos, which are typically made negligible by requiring good primary vertices and isolated leptons.  
Because of the non-prompt signature, dedicated trigger paths are needed to capture signal events. 
The ATLAS searches are performed in displaced dijet and lepton jets final states \cite{ATLAS-CONF-2014-041,arXiv:1409.0746}.
For a bench mark scenario, exclusion limits are placed on the $\pi_{v}$ proper decay length (see Fig.~\ref{fig:hvatlascms}(a)).
Exclusion limits are also set in the kinetic mixing parameter, $\epsilon$, and dark photon ($\gamma_{d}$) mass plane at 90\% CL (see Fig.~\ref{fig:hvatlascms}(b)).

\section{Summary}
The ATLAS and CMS collaborations search a variety of theoretical models beyond the SM.
Most of them are explored in more than one final state to account for a priori unknown couplings to SM particles.
No evidence of new exotic physics is found with the full $\sim$20 $fb^{-1}$ 8TeV Run I data.
Therefore, exclusion limits are set at $\sim$TeV scales depending on the models and parameters.
Only a subset of recent searches are presented in this document. 
Other search results can be found at the public web pages of the ATLAS and CMS collaborations \cite{atlaspub, cmspub}. 
LHC Run II at higher center-of-mass energy will start in 2015. 
This data will provide another opportunity to discover exotic phenomena, if there are any.


\begin{thebibliography}{99}
\bibitem{cmshiggs}
CMS Collaboration, Precise determination of the mass of the Higgs boson and studies of the compatibility of its couplings with the standard model, CMS-PAS-HIG-14-009
\end{thebibliography}
\end{document}